\documentstyle[12pt,epsf]{article}
\begin{document}

\title{ Hamilton-Jacobi theory for one dimensional autonomous systems using
parametric transformations }
\author{ G. Gonz\'alez \thanks{%
E-mail: gabriel${}_{-}$glez@yahoo.com}, \\
{\it Departamento de Matem\'aticas y F{\'\i}sica} \\
{\it I.T.E.S.O.}\\
{\it Perif\'erico Sur \# 8585 C.P. 45090} \\
{\it Guadalajara, Jalisco, M\'exico.}}
\date{}
\maketitle

\begin{abstract}
A necessary and sufficient condition for a parameter
transformation that leaves invariant the energy of a one
dimensional autonomous system is obtained. Using a parameter
transformation the Hamilton-Jacobi equation is solved by a
quadrature. An example of this approach is given.
\end{abstract}

\newpage

\section{Introduction}

It is well known that the problem of integrating Hamilton's
canonical equations of motion, which are expressed as a system of
ordinary differential equations of the first order, can be
replaced by an equivalent problem of finding a complete solution
of a single non linear partial differential equation of the first
order, called the Hamilton-Jacobi equation \cite{B1}. If we
succeed in finding a complete solution of the Hamilton-Jacobi
partial differential equation, the motion of the dynamical system
can be easily found directly from this complete solution by using
only elementary algebraic operations \cite{Vu}. \\
The Hamilton-Jacobi equation can be considered the
most elegant and powerful method known for finding the general
solution to the mechanical equations of motion and gives an
important physical example of the deep connection between first
order partial differential equations and first order ordinary
differential equations \cite{Vi}. \\
Each complete integral of the Hamilton-Jacobi equation gives rise
to a family of solutions of Hamilton's equations and is the
generating function of the canonical transformation which maps the
dynamical system to a trivial one with vanishing Hamiltonian. \\
Apart from its practical aspects, the Hamilton-Jacobi theory leads
to a geometric picture of dynamics relating the dynamics to wave
motion and has been the starting point for Schr\"{o}dinger to
state the wave equation in quantum mechanics \cite{Co}. \\
The applicability of using the Hamilton-Jacobi procedure depends
in our ability to solve a first order non linear partial
differential equation which can be very complicated to solve in
some cases \cite{GLo}. The main purpose of this article is to show
that, under certain conditions, it is possible to introduce a
parameter transformation which allow us to obtain a complete
integral for the Hamilton-Jacobi equation for one dimensional
autonomous systems without having to solve the partial
differential equation given by the Hamilton-Jacobi method.

\section{Parameter transformations and the Hamilton-Jacobi equation}

Newton's equation of motion for one dimensional autonomous systems can be
written as the following dynamical system
\begin{equation}
\frac{dx}{dt}=v,\qquad \frac{dv}{dt}=F(x,v),  \label{eq1}
\end{equation}%
where $x$ is the position of the particle, $v$ is the velocity and
$F(x,v)$ is the force divided by the mass of the particle. It is
well known that the Lagrangian $L(x,v)$ for a one dimensional
system always exist \cite{D} thus we may obtain (\ref{eq1}) from
the Euler-Lagrange equation \cite{B1}
\begin{equation}
\frac{d}{dt}\left( \frac{\partial L}{\partial v}\right) =\frac{\partial L}{%
\partial x},  \label{eq2}
\end{equation}%
and the associated constant of motion for (\ref{eq1}) from the Legendre
transformation \cite{B1}
\begin{equation}
K(x,v)=v\frac{\partial L}{\partial v}-L,  \label{eq3}
\end{equation}%
the numerical value of (\ref{eq3}) will be referred as the energy
of the system. Knowing the constant of motion of (\ref{eq1}) we
can obtain the associated Hamiltonian by expressing the constant
of motion in terms of the canonical variables $H(x,p)=K(x,v(x,p))$
and using this Hamiltonian we obtain the Hamitlon-Jacobi equation
\begin{equation}
H\left(x,\frac{\partial S}{\partial x}\right)+\frac{\partial
S}{\partial t}=0, \label{eq3a}
\end{equation}
where $S$ is known as Hamitlon's principal function \cite{B1}.\\
Let us now consider the following parameter transformation of the
form
\begin{equation}
\tau =\tau (t),\qquad \tilde{x}=x(t),  \label{eq4}
\end{equation}
which does not affect the position variable and it will be assumed
that the functions $\tau (t)$ are of class $C^{2}$ such that
$d\tau /dt>0$. The problem is to find a parameter transformation
which leaves invariant the energy of the system when subject to
this type of transformation. To do this let $L(x,v)$ and
$\tilde{L}(\tilde{x},\tilde{v})$ be the Lagrangians for
(\ref{eq1}) before and after the parameter transformation
respectively. Assume that $\partial ^{2}L/\partial v^{2}$ and
$\partial ^{2}\tilde{L}/\partial \tilde{v}^{2}$ do not vanish or
become infinite in some region $R$ of the dynamical space then we
have the following

\newtheorem{ein}{Theorem}
\begin{ein}
A necessary and sufficient condition for a parameter
transformation to leave invariant the energy of a one dimensional
autonomous system is that \\
\[ \frac{\partial^{2}L}{\partial v^2}\frac{dv}{dt}=\frac{\partial^{2}\tilde{L}}
{\partial \tilde{v}^2}\frac{d\tilde{v}}{d\tau}. \]
\end{ein}

\newtheorem{proof}{Proof}
\begin{proof}

The Euler-Lagrange equation for $L(x,v)$ may be written as
\[
\frac{\partial^{2}L}{\partial x\partial
v}v+\frac{\partial^{2}L}{\partial v^2}\frac{dv}{dt}=\frac{\partial
L }{\partial x}
\]
similarly the Euler-Lagrange equation for
$\tilde{L}(\tilde{x},\tilde{v})$ is given by
\[
\frac{\partial^{2}\tilde{L}}{\partial \tilde{x}\partial
\tilde{v}}\tilde{v}+\frac{\partial^{2}\tilde{L}}{\partial
\tilde{v}^2}\frac{d\tilde{v}}{d\tau}=\frac{\partial \tilde{L}
}{\partial \tilde{x}},
\]
by assumption of the theorem we have
\[
\frac{\partial L }{\partial x}-\frac{\partial^{2}L}{\partial
x\partial v}v=\frac{\partial \tilde{L} }{\partial
\tilde{x}}-\frac{\partial^{2}\tilde{L}}{\partial \tilde{x}\partial
\tilde{v}}\tilde{v}
\]
therefore
\[
\frac{\partial}{\partial x}\left(L-v\frac{\partial L }{\partial v}
\right)= \frac{\partial}{\partial
\tilde{x}}\left(\tilde{L}-\tilde{v}\frac{\partial \tilde{L}
}{\partial \tilde{v}} \right),
\]
which implies
\[
\frac{\partial K(x,v)}{\partial
x}=\frac{\partial\tilde{K}(\tilde{x},\tilde{v})}{\partial
\tilde{x}}.
\]
Taking the total time derivative of
$\tilde{K}(\tilde{x},\tilde{v})$ we obtain
\[
\frac{d\tilde{K}}{dt}=\frac{\partial\tilde{K}}{\partial
\tilde{x}}\frac{d\tilde{x}}{d\tau}\frac{d\tau}{dt}+\frac{\partial\tilde{K}}{\partial
\tilde{v}}\frac{d\tilde{v}}{d\tau}\frac{d\tau}{dt},
\]
using the fact that $\partial
\tilde{K}/\partial\tilde{v}=(\partial ^{2}\tilde{L}/\partial
\tilde{v}^{2})\tilde{v}$ then
\[
\frac{d\tilde{K}}{dt}=\frac{d\tilde{x}}{d\tau}\frac{d\tau}{dt}\left(\frac{\partial
\tilde{K}}{\partial\tilde{x}}+\frac{\partial^{2}\tilde{L}}
{\partial \tilde{v}^2}\frac{d\tilde{v}}{d\tau} \right)
\]
but $dx/dt=(d\tilde{x}/d\tau)(d\tau/dt)$, $\partial K/\partial
x=\partial\tilde{K}/\partial \tilde{x}$ and $\partial
^{2}L/\partial v^{2}(dv/dt)=\partial ^{2}\tilde{L}/\partial
\tilde{v}^{2}(d\tilde{v}/d\tau)$ then
\[
\frac{d\tilde{K}}{dt}=\frac{dK}{dt}
\]
which implies that $K(x,v)=\tilde{K}(\tilde{x},\tilde{v})$ and
completes the proof of the theorem.

\end{proof}

The main advantage of using this type of parameter transformation
is that neither the position or the energy of the system are
affected, using this fact and taking into account that we are
dealing with an autonomous system we can express the solution to
the Hamilton-Jacobi equation as \cite{B1}
\begin{equation}
S=W(x)-Et, \qquad \tilde{S}=\tilde{W}(x)-E\tau, \label{eq5}
\end{equation}

where $S$ and $\tilde{S}$ are Hamilton's principal functions for
(\ref{eq1}) before and after the parameter transformation
respectively and $W$ and $\tilde{W}$ represents the solution to a
nonlinear first order partial differential equation given by
\cite{B1}
\begin{equation}
H\left(x,\frac{dW}{dx}\right)=E, \qquad
\tilde{H}\left(x,\frac{d\tilde{W}}{dx}\right)=E, \label{eq6}
\end{equation}
which is usually known as Hamilton's characteristic function. If
we know how the generalized linear momentum transforms under
(\ref{eq4}) and recalling that
\begin{equation}
\frac{dW}{dx}=p, \qquad \frac{d\tilde{W}}{dx}=\tilde{p},
\label{eq7}
\end{equation}
we may express Hamilton's principal function for (\ref{eq1}) as
\begin{equation}
S=\int p(x,\tilde{p}(x,E))\,dx-Et. \label{eq8}
\end{equation}
Therefore, we obtain the solution for the Hamilton-Jacobi equation
by a quadrature. This way of solving the Hamilton-Jacobi equation
is useful to solve complicated problems as it would be shown in
the following example.
\section{Example}
Consider a relativistic particle of mass at rest $m$ under the
action of a constant force $\lambda>0$ and immersed in a medium
that exerts some type of friction which is proportional to the
square of the velocity. The classical equation of motion for this
system is given by
\begin{equation}
m\frac{dv}{dt}=(\lambda-\gamma v^2)(1-v^2/c^2)^{3/2}, \label{eq9}
\end{equation}
where $\gamma$ is a positive real parameter and $c$ represents the
speed of light. Writing (\ref{eq9}) at first order of
approximation in $v^2/c^2$ we have
\begin{equation}
m\frac{dv}{dt}=(\lambda-\gamma v^2)(1-\alpha^2 v^2), \label{eq10}
\end{equation}
where $\alpha^2=\frac{3}{2c^2}$ .The Lagrangian associated to this
system is \cite{G1}
\begin{equation}
L(x,v)=\frac{mv\tanh^{-1}(\alpha v)}{2\alpha}
e^{-2x(\lambda\alpha^2-\gamma)/m}-\frac{m\lambda}{2(\lambda\alpha^2-\gamma)}\left(
e^{-2x(\lambda\alpha^2-\gamma)/m} - 1\right), \label{eq11}
\end{equation}
and the Hamiltonian of this system is given by \cite{G1}
\begin{equation}
H(x,p)= \frac{m}{2}
e^{-2x(\lambda\alpha^2-\gamma)/m}\sum_{n=0}^{\infty}\alpha^{2n}
v^{2n+2}(x,p) +\frac{m\lambda}{2(\lambda\alpha^2-\gamma)}\left(
e^{-2x(\lambda\alpha^2-\gamma)/m} - 1\right), \label{eq12}
\end{equation}
where $v^{2n+2}(x,p)$ is given by
\begin{equation}
v^{2n+2}(x,p)= \left(\frac{p}{m}e^{-2\gamma
x/m}\frac{2n+1}{(n+1)!}\left(\frac{ 2\lambda
x}{m}\right)^{n}\right)^{(2n+2)/(2n+1)}. \label{eq13}
\end{equation}
The Hamiltonian (\ref{eq12}) is valid for the case $|\alpha v|<1$
and has physical meaning only when $p>0$ and $x>0$. Once knowing
the Hamiltonian we can obtain the Hamilton-Jacobi equation for the
system
\begin{eqnarray}
\frac{m}{2}
e^{-2x(\lambda\alpha^2-\gamma)/m}\sum_{n=0}^{\infty}\alpha^{2n}\left(\frac{\partial
S }{\partial x}e^{-2\gamma x/m}\frac{2n+1}{m(n+1)!}\left(\frac{
2\lambda x}{m}\right)^{n}\right)^{(2n+2)/(2n+1)} + \nonumber\\
\frac{m\lambda}{2(\lambda\alpha^2-\gamma)}\left(
e^{-2x(\lambda\alpha^2-\gamma)/m} - 1\right) +\frac{\partial S
}{\partial t}=0,\label{eq14}
\end{eqnarray}
therefore one has to solve (\ref{eq14}) to obtain Hamilton's
principal function, which at first sight may seem like a
formidable task, but it can be done applying the approach
described in the last section.\\
Consider the following parameter transformation
\begin{equation}
\tau=\int_{0}^{t} \sqrt{1-\alpha^{2}v^{2}}\,dt, \qquad
\tilde{x}=x, \label{eq15}
\end{equation}
therefore equation (\ref{eq10}) transforms into
\begin{equation}
m\frac{d\tilde{v}}{d\tau}=\lambda+\tilde{\gamma}\tilde{v}^2,
\label{eq16}
\end{equation}
where $\tilde{\gamma}=\lambda\alpha^2-\gamma$ and the
transformation equations between one set of dynamical variables
$(x,v)$ to the other set of dynamical variables
$(\tilde{x},\tilde{v})$ are given by
\begin{eqnarray}
{x=\tilde{x} \atop
v=\frac{\tilde{v}}{\sqrt{1+\alpha^2\tilde{v}^{2}}}} \qquad
{\tilde{x}=x \atop \tilde{v}=\frac{v}{\sqrt{1-\alpha^2v^{2}}}}
\label{eq17}
\end{eqnarray}
The Lagrangian and the Hamiltonian for (\ref{eq16}) can be
obtained, and are given by \cite{GL2}
\begin{equation}
\tilde{L}(x,\tilde{v})=\frac{m}{2}\tilde{v}^2
e^{-2\tilde{\gamma}x/m}-\frac{m\lambda}{2\tilde{\gamma}}\left(
e^{-2\tilde{\gamma}x/m} - 1\right), \label{eq18}
\end{equation}
\begin{equation}
\tilde{H}(x,\tilde{p})=\frac{\tilde{p}^2}{2m}e^{2\tilde{\gamma}x/m}+
\frac{m\lambda}{2\tilde{\gamma}}\left(e^{-2\tilde{\gamma}x/m} -
1\right),  \label{eq19}
\end{equation}
where we have dropped the tilde for the position variable for the
sake of simplicity. \\
Using (\ref{eq9}),(\ref{eq11}), (\ref{eq16}) and (\ref{eq18}) it
is easy to convince oneself that theorem $1$ is fulfill, therefore
the energy is invariant under transformation (\ref{eq15}). What we
need now is to find how the generalized linear momentum transforms
under (\ref{eq15}), to do that we express
$\tilde{p}=m\tilde{v}exp(-2\tilde{\gamma}x/m)$ in terms of $v$
using (\ref{eq17}), which gives
\begin{equation}
\tilde{p}=m\tilde{v}e^{-2\tilde{\gamma}x/m}=\frac{mve^{-2\tilde{\gamma}x/m}}
{\sqrt{1-\alpha^2v^{2}}}=me^{-2\tilde{\gamma}x/m}\sum_{n=0}^\infty
\frac{(2n)!\alpha^{2n}v^{2n+1}}{2^{2n}(n!)^2}, \label{eq20}
\end{equation}
where we have used the fact that $|\alpha v|<1$ in the last step.
Substituting (\ref{eq13}) into (\ref{eq20}) we have the way the
generalized linear momentum transforms under (\ref{eq15})
\begin{equation}
\tilde{p}=pe^{-2\lambda\alpha^2 x/m}\sum_{n=0}^\infty
\frac{(2n+1)(2n)!}{2^{n}(n+1)!(n!)^2}\left(\frac{\lambda\alpha^2
x}{m} \right)^{n}=pe^{-\lambda\alpha^2 x/m}\left(
I_{0}(\lambda\alpha^2 x/m)+I_{1}(\lambda\alpha^2 x/m)
\right),\label{eq21}
\end{equation}
where $I_{n}(z)$ is the modified Bessel function of the first kind
\cite{A} . Using the fact that
\begin{equation}
\tilde{p}=e^{-\tilde{\gamma}x/m}\sqrt{2mE-\frac{m^2\lambda}{\tilde{\gamma}}
\left(e^{-2\tilde{\gamma}x/m}-1\right)}, \label{eq22}
\end{equation}
and substituting (\ref{eq22}) into (\ref{eq21}) we have
\begin{equation}
p(x,\tilde{p}(x,E))= \frac{e^{\gamma
x/m}\sqrt{2mE-\frac{m^2\lambda}{\lambda\alpha^2-\gamma}
\left(e^{-2(\lambda\alpha^2-\gamma)x/m}-1\right)}}{I_{0}(\lambda\alpha^2
x/m)+I_{1}(\lambda\alpha^2 x/m)},\label{eq23}
\end{equation}
substituting (\ref{eq23}) into (\ref{eq8}) we obtain Hamilton's
principal function for $x>0$
\begin{equation}
S=\int \frac{e^{\gamma
x/m}\sqrt{2mE-\frac{m^2\lambda}{\lambda\alpha^2-\gamma}
\left(e^{-2(\lambda\alpha^2-\gamma)x/m}-1\right)}}{I_{0}(\lambda\alpha^2
x/m)+I_{1}(\lambda\alpha^2 x/m)}\,dx-Et.\label{eq24}
\end{equation}
All the expressions derived in this paper have the right limit when
$\gamma\rightarrow 0$ and $\alpha \rightarrow 0$.  \\

\section{Conclusions}
A necessary and sufficient condition for a parameter
transformation that leaves invariant the energy of a one
dimensional autonomous system was obtained. A method for solving
the Hamilton-Jacobi equation using a parameter transformation was
deduced. All the expressions obtained in this paper converge to
the conservative case
when the dissipation parameter goes to zero. \\

~~~~~\\

\newpage


\end{document}